\documentstyle[twocolumn,prl,aps,eqsecnum]{revtex}
\begin{document}
\draft
\title{Characterization of One-Dimensional Luttinger 
Liquids in Terms of Fractional Exclusion Statistics}
\author{ Yong-Shi Wu $^1$ and Yue Yu $^2$ and Huan-Xiong Yang$^{3,2}$}
\address{1. Department of Physics, University of Utah, 
Salt Lake City, UT 84112, U.S.A.}
\address{2. Institute of Theoretical Physics, Chinese 
Academy of Sciences, Beijing 100080, P. R. China}
\address{3. Department of Physics, Zhejiang University, 
Hangzhou, 310027, P. R. China} 
\maketitle
\begin{abstract}
We develop a bosonization approach to study the 
low temperature properties of one-dimensional 
gas of particles obeying fractional exclusion 
statistics (FES). It is shown that such ideal 
gas reproduces the low-energy excitations and 
asymptotic exponents of a one-component 
Luttinger liquid (with no internal degrees 
of freedom). The bosonized effective theory at 
low energy (or temperature) is identified to 
a $c=1$ conformal field theory (CFT) with 
compactified radius determined by the statistics 
parameter $\lambda$. Moreover, this CFT can be 
put into a form of the harmonic fluid description 
for Luttinger liquids, with the Haldane controlling 
parameter identified with the statistics parameter 
(of quasi-particle excitations). Thus we propose 
to use the latter to characterize the fixed points 
of 1-d Luttinger liquids. Such a characterization 
is further shown to be valid for generalized ideal 
gas of particles with mutual statistics in momentum 
space and for non-ideal gas with Luttinger-type 
interactions: In either case, the low temperature 
behavior is controlled by an effective statistics 
varying in a fixed-point line.
\end{abstract}



\section{Introduction}

It is well-known that the Landau theory of Fermi 
liquids fails to describe most of one-dimensional 
(1-d) interacting many-body systems. To provide
a substitute, Haldane proposed, years ago, the 
concept of the Luttinger liquid \cite{Hald1}, 
defined by a set of low-lying excitations and 
critical exponents of the asymptotic correlation 
functions. Like Fermi liquids, there is a 
(pseudo-)Fermi surface for the quasiparticle-like 
excitations in Luttinger liquids, so that the 
classification of low-lying excitations is similar 
to that in Fermi liquids. However, the exponents 
of the asymptotic correlation functions (at low 
temperature) are distinct from those for Fermi 
liquid theory. For one-component systems (without 
internal degrees of freedom), the low-energy or 
low-temperature behavior of a Luttinger liquid 
is controlled by a single parameter, the Haldane 
controlling parameter. It controls not only all 
exponents, but also the velocity ratios between 
different types of elementary excitations. The 
Fermi liquid theory is a special case of the 
Luttinger liquids with the Haldane parameter 
$\lambda=1$. 

In recent years, the failure of Landau's theory 
of Fermi liquids to describe several newly 
discovered strongly correlated electron systems 
have revived the interests in the theory of 
Luttinger liquids. Among other questions, 
compared with Fermi liquids, one would like 
very much to know the answer to the following 
questions: 
\begin{itemize}
\item What is the physical meaning of Haldane's 
controlling parameter? Or more precisely, how to 
use physical properties of low-lying excitations 
to characterize the concept of Luttinger liquids?
\item Does Haldane's theory of Luttinger liquids
possess universality in one dimension, just like 
Landau's theory of Fermi liquids in three 
dimensions? Or equivalently, in terms of modern 
language of renormalization group, does the 
Luttinger liquids describe the infrared (or 
low-energy) fixed points in 1-d systems?
\item In what directions could one expect to 
go for generalizing the concept of Luttinger 
liquids to higher than one dimensions? 
\end{itemize}
In short, a characterization of 1-d Luttinger 
liquids, other than using a bunch of excitations
and exponents, is in demand for gaining more 
insights and looking for possible generalization.   

To achieve this, let us recall what motivated 
Landau's concept of Fermi liquids, which is 
known to describe an infrared fixed point (or 
a universality class) of interacting electron 
systems. The basic idea behind it is based on the 
following {\it organizing principle} for 
interacting many-body systems: {\it At low 
temperature, the low-lying excited states of an 
interacting many-body system above a stable ground 
state can be viewed as consisting of weakly coupled 
elementary excitations.} Here "weakly coupled" only 
means that the total energy can be written as a sum 
of single-particle (dressed) energies, while the 
dispersion of the dressed energy may well depend on 
the total particle number, a signal of remnant 
interactions between the quasiparticles. According 
to Landau, the ground state and the low-lying excited 
states of a Fermi liquid are approximately, to a good
accuracy at sufficiently low temperature, described 
by those of an ideal Fermi gas with dressed energy 
for the quasiparticles.

We note the significant role played by the ideal 
Fermi gas distribution (with dressed energy) in 
this description of Fermi liquids. Actually it is
the ideal Fermi gas that gives a characterization 
to the Fermi liquid fixed point, and a meaning to
the universality of the concept of Fermi liquids.
This inspires us to try to give a characterization 
of the 1-d Luttinger liquids along a similar line 
of thoughts, namely using a properly generalized 
concept of {\it exclusion statistics}, of which 
a special case is the usual Fermi statistics.
Because the concept of quantum statistics in 
statistical mechanics is independent of the 
dimensionality of a system, a characterization 
of 1-d infrared fixed points using statistics, if
successful, would shed light on how to generalize 
to higher dimensional non-Fermi liquids.
 
Fortunately, a generalization as such has been 
available recently, under the name of fractional 
exclusion statistics (FES). It is based on a new 
combinatoric rule for the many-body state counting 
\cite{Hald2,Wu}, which is essentially an abstraction 
and generalization of Yang-Yang's state counting 
\cite{YangYang,BerWu} in 1-d soluble many-body 
models. FES has been shown to be applicable to 
elementary excitations in a number of exactly 
solvable models for strongly correlated systems 
\cite{Hald2,Wu,BerWu,NaWil,Ha,Hatsu,WuYu},
anyons in the lowest Landau level \cite{Ouvry,Wu}, 
and quasiparticle excitations in the fractional 
quantum Hall effect \cite{Hald2,Wu,Ha,HKWY}. The 
thermodynamics of the so-called generalized ideal 
gas (GIG) associated with FES have been studied 
\cite{Wu} in a general framework. 

Inspired by these results, the thoughts along the 
lines indicated in the above paragraphs have led 
two of present authors \cite{WuYu1} to propose that 
at least {\it for some strongly correlated systems 
or non-Fermi liquids, their low-energy or 
low-temperature fixed point may be described by 
a GIG associated with FES}, similar to the way 
that of the Fermi liquid fixed point by the ideal 
Fermi gas \cite{RG}. As a testimony to this 
proposition, a sketchy proof was given in that 
short letter \cite{WuYu1} that the low-$T$ critical 
properties of the 1-d Luttinger liquids are exactly 
reproduced by those of 1D ideal excluson gas(IEG), 
if one identifies the Haldane parameter of the 
former with the statistics parameter $\lambda$ of 
the latter. (We call the particles obeying the 
FES without mutual statistics {\it exclusons}). 
Threfore {\it IEG can be used to describe the 
fixed points of the Luttinger liquids}. In this 
paper, we will present our results obtained 
in \cite{WuYu1} in details, much of which was 
not published before. 

A main tool we use in this study is bosonization
of the 1-d excluson systems at low $T$, {\it \`a 
la} Tomonaga \cite{Tomonaga} and Mattis and 
Lieb \cite{ML}. To bosonize an IEG system is a 
little bit tricky, because at low temperature the 
linearized dispersion of dressed energy versus
pseudo-momentum has different slope outside and 
inside the pseudo-Fermi sea: There is `refraction' 
at both pseudo-Fermi points. In spite of this, 
we still manage to construct well-defined density 
fluctuation operators that obey the $U(1)$ current
algebras and physically describe free phonons. Then, 
the Tomonaga-Mattis-Lieb bosonization applies, 
resulting a bosonized effective field theory, in 
agreement with Haldane's harmonic fluid description 
of the Luttinger liquid \cite{Hald3}. Then the 
asymptotic correlation functions and their 
exponents can be systematically calculated. In 
this way, the critical properties of IEG reproduce 
those of the Luttinger liquids. 

An important consequence of our bosonization is 
that the low energy behavior of IEG is controlled 
by an {\it orbifold} conformal field theory (CFT) 
with central charge $c=1$ and compactified radius 
\cite{CFT} $R=\sqrt{1/\lambda}$. This variant of 
$c=1$ CFT is {\it not} the ordinary $c=1$ CFT 
compactified on a circle $S^1$, rather it is 
compactified on an orbifold $S^1/Z_2$, which is 
topologically an interval \cite{CFT}. The 
differences arise due to different selection rules 
for vertex operators, that constrain quantum 
numbers of possible quasiparticle excitations in 
the system. In the usual literature this difference 
quite often is overlooked. Only within the orbifold 
CFT the IEG with statistics parameter $\lambda=1$ 
recovers ideal Fermi gas, as it should be. Also the 
two classes of $c=1$ CFT's have different duality 
relation; only the one in orbifold CFT reproduces 
the known particle-hole duality in IEG, $\lambda 
\leftrightarrow 1/\lambda$, as given in 
\cite{BerWu,NaWil}. (For the details and more
elaboration, see below. We note \cite{WuYu}
that a similar situation happens for the 
Calogero-Sutherland (C-S) model\cite{Cal,Suth}: 
The low-energy effective field theory for the 
bosonic and fermionic C-S models belongs to, 
respectively, the above-mentioned two classes 
of $c=1$ CFT.) 

The fact that the low-$T$ behavior of IEG is 
controlled by a {\it conformally invariant} 
theory is significant, implying that indeed 
IEG provides a characterization of infrared 
fixed points, having the conformal invariance 
as required by renormalization group. We have 
also studied the effects of mutual statistics 
between different pseudomomenta and of the 
Luttinger-type (density-density) interactions 
among exclusons. In either case, the low-$T$ 
behavior is controlled by an effective 
statistics $\lambda_{eff}$ for excitations 
near the Fermi points, the same way as 
$\lambda$ in the case of IEG. In one 
dimension both the momentum-independent 
part of interactions and change in chemical 
potential $\mu$ are {\it relevant} 
perturbations \cite{RG,Schulz}, leading to 
a continuous shift in the fixed-point line 
parameterized by $\lambda$. All these will be 
explained in details in the present paper.

To make this paper self-contained, we devote 
the next two sections, Sec. II and Sec. III, 
to reviewing the Luttinger liquid theory and 
the GIG associated with FES, respectively. In 
Sec. IV, we discuss the low-energy behavior 
of the IEG system and achieve its bosonization. 
In Sec. V, the generalization to the GIG with 
mutual statistics as well as the non-ideal gas 
with FES are provided. The last section is 
dedicated to conclusions and discussions.

\section{Luttinger liquid}

The Luttinger liquid, which describes a very
large class of one-dimensional interacting 
many-body systems, is introduced because of 
the infrared divergence of certain vertices 
in the Fermi liquid description of the 1-d
systems. Some pioneering works have been done 
in the Luttinger model before the Luttinger 
liquid concept \cite{Lutt,ML}. The model has 
been exactly solved by using the bosonization 
technique \cite{ML}. Haldane \cite{Hald1}
re-solved the model with the following 
important observations:

\noindent{
(i) Besides a linearized spectrum of non-zero 
mode excitations, i.e., the density fluctuations 
(sound waves), there are two kinds of zero mode 
excitations, single-particle excitations by 
adding extra particles to the system and 
persistent currents by making Galileo boosts.
}

\noindent{
(ii) There is a fundamental relation among the 
velocities of these three types of excitations
}
\begin{equation}
v_s = \sqrt{v_N \, v_J}, 
\label{VeRe}
\end{equation}    
where $v_s$ is the sound velocity, $v_J$ the 
current velocity and $v_N$ a velocity related  
to the change in particle number. The velocity 
ratios define a controlling parameter, 
$e^{-2\varphi}$, by 
\begin{equation}
v_N=v_se^{-2\varphi},\hspace{.2in}
v_J=v_se^{2\varphi}.
\label{VeRePh}
\end{equation}

\noindent{
(iii) The above defined controlling parameter 
measures the essential renormalized coupling 
constant, and is the unique parameter that 
determines the exponents of power-law decay 
in the zero-temperature correlation functions.
}

Based on these observations, Haldane defined 
the Luttinger liquids as 1-d systems that have
similar behavior (i)-(iii) at low temperature 
just like the Luttinger model. In this way the 
Luttinger liquids are characterized through 
their excitations and the exponents of the 
asymptotic correlation functions. 

To be more precise, recall that the Luttinger 
model describes a one-dimensional interacting 
fermion system with the Hamiltonian
\begin{equation}
H=\int dx |\nabla\psi|^2+\frac{1}{2}
\int\int dx dy V(x-y)\rho(x)\rho(y).
\label{Hm}
\end{equation}
In the low energy limit, the Hamiltonian 
(\ref{Hm}) can be bosonized as
\begin{equation}
H=v_s\sum_{q} |q| b_q^\dagger b_q
+\frac{1}{2}(\pi/L)(v_N\, M^2+v_J\,J^2),
\label{HmB}
\end{equation}
where $b_q$ are the standard boson annihilation 
operators, and $M$ and $J$ the operators 
corresponding to adding extra particle and 
boosting persistent currents, whose eigenvalues 
obey the following selection rule,
\begin{equation}
(-1)^J=(-1)^M.
\label{select}
\end{equation}
The total momentum of the system also has 
a bosonized form
\begin{equation}
P=[k_F+(\pi/)M]J+\sum_qqb^\dagger_qb_q,
\label{MmB}
\end{equation}
with $k_F$ being the Fermi momentum. 

Eqs.(\ref{VeRe},\ref{VeRePh},\ref{HmB}-\ref{MmB}) 
turned to be universally valid for the description 
of the low-energy properties of gapless 
interacting one-dimensional spinless fermion 
systems even for those not exactly soluble with 
a conserved current $J$. This universality class 
is named as the Luttinger liquid by Haldane 
\cite{Hald1}. The Luttinger liquid has a 
model-independent representation, namely the 
harmonic fluid description \cite{Hald3}, which 
is convenient for calculating the correlation 
functions. The results of the harmonic fluid
representation are listed in the Appendix, for 
later use to be compared with our bosonization 
theory of the IEG.

The Haldane theory of Luttinger liquids is 
based on the significant observation that 
the low-$T$ behavior of the Luttinger model 
is universal. Naturally arises the question: 
Why is it so? In this paper we intend to 
answer this question by pointing out a 
profound coincidence of the low-$T$ behavior 
of the Luttinger model and that of ideal 
excluson gas (IEG), i.e., ideal gas of particles 
obeying fractional exclusion statistics: The
universality of the former is due to that of 
the latter. 

\section{Generalized ideal gas}

In quantum mechanics, there are two ways to 
define the statistics of particles. One is in 
terms of the symmetry of the many-body wave 
function under particle exchange. The other 
is based on the state counting. Here we are 
interested in the latter definition. As is
well-known, bosons and fermions have different 
countings for many-body states, or different 
statistical weights $W$: The number of 
quantum states of $N$ particles occupying a 
group of $G$ states is, for bosons and 
fermions respectively, given by
\begin{equation}
W_{b}= {(G+N-1)! \over N!~ (G-1)!}~,~~~
{\rm{or}}~~ W_{f}= {G! \over N!~ (G-N)!}~.    
\label{2}
\end{equation}
A simple interpolation between bosons and 
fermions is given by \cite{Hald2,Wu}
\begin{equation}
 W = {[G+(N-1)(1-\lambda)]!
\over N!~ [G-\lambda N-(1-\lambda)]!}~,  
\label{2'}
\end{equation}
with $\lambda=0$ corresponding to bosons 
and $\lambda=1$ to fermions. The physical 
meaning of this equation is the following: 
By assumption, the statistical weight remains 
to be {\it a single combinatoric number}, so 
one can count the states by thinking of the 
particles {\it effectively either as bosons 
or as fermions}, with the effective number 
of available single-particle states being 
{\it linearly dependent on the particle 
number}:
\begin{equation}
G_{eff}^{(b)} = G - \lambda (N-1), \;\;\;
{\rm or}\;\;\;\; G_{eff}^{(f)} 
= G - (1-\lambda) (N-1).    
\label{3}
\end{equation}
Obviously, for genuine bosons (or fermions), 
$G_{eff}^{(b)}\;$ (or $G_{eff}^{(f)}\;$) is 
independent of the particle number. In all 
other cases, either of the two $G_{eff}$ is 
linearly dependent on the particle number. 
This is the defining feature of the FES. The 
statistics parameter $\lambda$ tells us, on 
the average, how many single-particle states 
that a particle can exclude others to occupy. 
A proper understanding of this has been 
discussed in \cite{WYS}. Thus, the expression 
(\ref{2'}) for the statistical weight, $W$, 
formulates a generalized Pauli exclusion 
principle, as first recognized by Haldane 
\cite{Hald2}.

It is easy to generalize this state
counting to more than one species,
labeled by the index $i$:
\begin{equation}
W = {\prod}_i ~ { [G_i + N_{i}-1 -
\sum_j \lambda_{ij}(N_j-\delta_{ij})]!
\over (N_i)!~ [G_i - 1- \sum_j \lambda_{ij}
(N_j-\delta_{ij})]! }~.   
\label{4}
\end{equation}
Here $G_i$ is the number of states when the 
system consists of only a single particle
of species $i$. By definition, the diagonal 
$\lambda_{ii}$ is the ``self-exclusion'' 
statistics of species $i$, while the 
non-diagonal $\lambda_{ij}$ (for $i\neq j$) 
is the mutual-exclusion statistics. Note 
that $\lambda_{ij}$, which Haldane 
\cite{Hald2} called {\it statistical 
interactions}, may be {\it asymmetric} in 
$i$ and $j$. The interpretation is similar 
to that of the one-species case: The number 
of available single-particle states for 
species $i$, in the presence of other 
particles, is again linearly dependent on 
particle numbers of all species:
\begin{eqnarray}
&&G^{(b)}_{eff,i}= G_i -\sum_{j}
\lambda_{ij} (N_j - \delta_{ji}),\nonumber\\    
&& {\rm or}\nonumber\\
&&G^{(f)}_{eff,i}= G^{(b)}_{eff,i}
+ N_i -1.    \label{5}
\end{eqnarray}

The definition (\ref{2'}) or (\ref{4}) starts 
with a postulated form for the statistical 
weight, and thus is more direct and convenient 
for the purpose of formulating quantum 
statistical mechanics. One of us \cite{Wu} 
has first formulated the quantum statistical 
mechanics by proposing the notion of generalized
ideal gas(GIG): A GIG satisfies the following 
two conditions: (i) The total energy (eigenvalue) 
is always of the form of a simple sum, in which 
the $i$-th term is linear in the particle number 
$N_{i}$:
\begin{equation}
 E=\sum_{i} N_{i} \varepsilon^0_{i},     
\label{7}
\end{equation}
with $\varepsilon^0_{i}$ identified as 
the energy of a particle of species $i$; 
(ii) The state-counting (\ref{4}) for 
statistical weight $W$ is applicable. 
When there are no statistical interactions
(i.e., $\lambda_{ij}=0$ for $i\neq j$), we
have the usual ideal gas, which we call
as IEG. 
 
With the assumptions (\ref{7}) and (\ref{4}), the 
thermodynamics of a GIG can be worked out by the 
usual techniques in statistical mechanics. 
Consider a grand canonical ensemble at temperature 
$T$ and with chemical potential $\mu_{i}$ for 
species $i$, whose partition function is given by
\begin{equation}
Z= \sum_{\{N_{i}\}} W(\{N_{i}\})~
\exp \{\sum_{i} N_{i} (\mu_{i} 
-\varepsilon^0_{i})/T \}~.    
\label{8}
\end{equation}
As usual, we expect that for very large $N_{i}$, 
the summation has a very sharp peak around the set 
of most-probable (or mean) particle numbers 
$\{\bar{N}_{i}\}$. Using the Stirling formula,
introducing the average ``occupation number
per state'' defined by 
$n_{i} \equiv \bar{N}_{i}/ G_{i}$, 
and maximizing
\begin{equation}
{\partial \over \partial n_{i}}\,
\bigl[ \ln W + \sum_{i}
G_{i} n_{i}\,(\mu_{i} -
\varepsilon^0_{i})/T \bigr] =0~,   
\label{9}
\end{equation}
one obtains the equations that determine the 
most-probable distribution of $n_{i}$ 
\begin{equation}
\sum_{j} (\delta_{ij}w_j +g_{ij}) n_j = 1~,  
\label{10}
\end{equation}
with $g_{ij}\equiv \lambda_{ij} G_{j}/G_{i}$, 
and $w_i$ being determined by the functional
equations
\begin{equation}
(1+w_i)  \prod_{j} \Bigl({w_j
\over 1+w_j}\Bigr)^{\lambda_{ji}}
= e^{(\varepsilon^0_i-\mu_{i})/T}.    
\label{11}
\end{equation}

The thermodynamic potential $\Omega=-T \ln Z$ 
and the entropy $S$ are then given by
\begin{eqnarray}
\Omega &\equiv &- PV = -T \sum_i G_i
\log {1+ n_i - \sum_j g_{ij} n_j
\over 1- \sum_j g_{ij} n_j}~\nonumber\\
&=&-T \sum_i G_i\ln (1+w_i^{-1}); \label{12}
\end{eqnarray}
\begin{eqnarray}
S&=& \sum_i G_i
\Bigl\{ n_i {\varepsilon^0_i - \mu_{i} \over T} +
\ln {1+ n_i - \sum_j g_{ij} n_j
\over 1- \sum_j g_{ij} n_j } \Bigr\}\nonumber\\
&=&\sum_i G_i\Bigl\{ n_i {\varepsilon^0_i 
- \mu_{i} \over T} + \ln (1+w_i^{-1}) \Bigr\} .    
\label{13}
\end{eqnarray}
Other thermodynamic functions follow
straightforwardly. As usual, one can 
easily verify that the fluctuations, 
$({\overline{{N_{i}}^{2}}}-{\bar{N_{i}}}^{2})/
{\bar{N_{i}}}^{2}$, of the occupation numbers
are negligible, which justifies the validity 
of the above approach.

\section{Bosonization of 1-d ideal excluson gas} 

Let us first consider the simplest case, the 
1-d IEG without internal degrees of freedom. 
We expect to obtain a continuous interpolation
between the usual ideal Bose and ideal Fermi gas. 
Moreover, we want to show that the low-energy 
behavior of the IEG reproduces that of the 
Luttinger liquid and, therefore, provides a 
better characterization of the infrared fixed 
points associated with the Luttinger liquid.

\subsection{Ideal Excluson Gas}
  
Consider a GIG of $N$ particles on a ring with 
size $L$. Single-particle states are labeled 
by pseudo-momenta $k_i$. The total energy and 
momentum are given by 
\begin{equation}
E=\sum k^2_i, \hspace{.2in}P=\sum k_i.
\end{equation}
According to (\ref{5}), in the thermodynamic 
limit the hole density, $\rho_a (k,T)$, (or 
the density of available single-particle states) 
is {\it linearly} dependent on the particle 
density, $\rho(k,T)$. By definition, the 
statistics interaction matrix is given by
\begin{equation}
\lambda(k_i,k_j)=
-\frac{\Delta\rho_a(k_i)}{\Delta\rho(k_j)}.
\end{equation}
Or in the thermodynamic limit, one has
  \begin{equation}
\lambda(k,k')= - \, \delta\rho_a(k)/ 
\delta\rho(k').
\label{stat} 
  \end{equation}
The system is called an IEG of statistics 
$\lambda$ (with {\it no mutual statistics} 
between different momenta), if 
\begin{equation}
\lambda(k_i,k_j)=\lambda\, \delta(k_i-k_j),
\end{equation} 
or (\ref{5}) reads
\begin{equation}
\rho(k_j)=\frac{1}{2\pi}
+\frac{1}{L}(1-\lambda)\sum_{i\not=j}
\delta(k_j-k_i)\rho(k_i)\Delta k,\label{rhoD}
\end{equation}
which, in the thermodynamic limit, can be 
simply written as \cite{BerWu}
\begin{equation} 
\rho_a(k,T)+\lambda\rho(k,T)=\rho_{0}(k,T),
\label{RHOT}
\end{equation} 
where $\rho_0(k)\equiv 1/2\pi$ is the bare 
density of single-particle states. Thus, 
$\lambda=1$ corresponds to fermions, and 
$\lambda=0$ to bosons. The thermodynamic 
potential, now reads, in terms of (\ref{12})
 \begin{equation}
\Omega=-\frac{T}{2\pi}\int_{-\infty}^\infty dk 
\,\ln(1+w(k,T)^{-1}),
\label{Omega}
\end{equation}
with the function $w(k,T)\equiv 
\rho_{a}(k)/\rho(k)$ 
satisfying an algebraic equation,
\begin{equation}
w(k,T)^{\lambda} [1+w(k,T)]^{1-\lambda}
=e^{(k^2-\mu)/T}. 
\label{forW}
\end{equation}

Firstly, we consider the ground state, in 
which the particles are distributed in a 
finite and origin-symmetric interval in the 
pseudo-momentum space. The (pseudo-)Fermi 
momentum is defined by 
\begin{equation}
k_F^2=\mu
\end{equation}
and its value is fixed by the average 
particle density 
$\bar{d}_0=N_0/L$ in the ground state, 
\begin{equation}
\int_{-k_F}^{k_F} dk\rho(k)=\bar{d}_0.
\end{equation}
Because holes are absent in the ground 
state, the particle density in the ground 
state is easily obtained from
(\ref{RHOT}),
\begin{equation}
\rho(k)=\Biggl\{\begin{array}{ll}
\displaystyle{1\over2\pi\lambda},
&{\rm  for}~~ |k|<k_{F};\\
0,&{\rm for}~~ |k|>k_{F}.
\end{array}\label{RHOZ}
\end{equation} 
Hence, one has
\begin{equation}
k_F=\pi\lambda\bar{d}_0,\hspace{.2in} 
\mu=(\pi\lambda\bar{d}_0)^2.
\end{equation}
Then the ground state energy and momentum 
are given by
\begin{eqnarray}
&&\frac{E_0}{L}=\int_{-k_F}^{k_F}dk \rho(k)k^2
={1\over 3}\pi^2\lambda^2 
\bar{d}_{0}^{3},\nonumber\\
&&P_{0}=\int_{-k_F}^{k_F}dk \rho(k)k=0.
\end{eqnarray}

Now let us examine possible excitations in an 
IEG. First there are density fluctuations due 
to particle-hole excitations, i.e., sound 
waves with velocity (see the next subsection)
\begin{equation}
v_{s} =v_{F}\equiv 2k_{F}.
\label{VSF}
\end{equation}
Besides, by adding extra $M$ particles to the 
ground state, one can create particle excitations, 
and by Galileo boosts a persistent current. We 
observe that the velocities of these three 
classes of elementary excitations in IEG also 
satisfy the fundamental relation (\ref{VeRe}). 
Indeed, shifting $N_0$ to $N=N_0+M$, the change 
in the ground state energy is 
\begin{eqnarray}
\delta_M E_0&=&{1\over 3}\pi^2\lambda^2 
(N/L)^{3}-{1\over 3}\pi^2\lambda^2 (N_0/L)^{3}
\nonumber\\
&=&\pi^2\bar{d}_0^2 M+\pi(\lambda k_F)M^2
+O(M^3/L^3),
\end{eqnarray}
while a persistent current, created by the 
boost of the Fermi sea $k\to k+\pi J/L$,
leads to the energy shift 
\begin{eqnarray}
\delta_JE_0 &=&\int_{-k_F+\pi J/L}^{k_F+\pi J/L}dk 
\rho(k)k^2-\int_{-k_F}^{k_F}dk \rho(k)k^2\nonumber\\
&=&\pi(k_F/\lambda)J^2.
\end{eqnarray}
Therefore the total change in energy,
due to charge and current excitations, is
\begin{equation} 
\delta E_0-\mu M=
\frac{\pi}{2L}v_F (\lambda M^2
+\lambda^{-1} J^2).
\label{EMC}
\end{equation}
The total momentum change due to the current 
excitations is
\begin{equation}
\delta P_0=\sum_k \frac{\pi J}{L}
=\pi(\bar{d}_0+\frac{M}{L})J.
\label{MC}
\end{equation}
If we denote the variation in free energy as 
$\delta F_0=\delta E_0-\mu M$, and identify 
$\lambda$ as the controlling parameter 
$e^{-2\varphi}$ in the Luttinger liquid 
theory, (\ref{EMC}) just recuperates the 
zero-mode contributions \cite{comm0} in 
(\ref{HmB}). Comparing (\ref{EMC}) with
(\ref{HmB}) we identify the velocities
$v_N$ and $v_J$ to be
 \begin{equation}
v_N=v_F\lambda, \qquad v_J=v_{F}/\lambda, 
\label{Velo}
\end{equation}
Then we see the velocity relation (\ref{VeRe}), 
i.e., $v_{s} = \sqrt{v_{N}\,v_{J}}$, that 
Haldane used to characterize the Luttinger 
liquids, is satisfied in IEG. The selection 
rule (\ref{select}) also holds for the IEG, 
since the system should correspond to the 
ideal Fermi gas if $\lambda=1$.

Encouraged by this relationship between the 
IEG and Luttinger liquids, we want to calculate 
the critical exponents of IEG to see whether
they reproduce those of the Luttinger liquids. 
This motivates to develop a bosonization 
for the density fluctuations in IEG.

\subsection{Low Energy Limit and Bosonization}  

Following Yang and Yang\cite{YangYang,Suth}, 
we introduce the dressed energy 
$\epsilon (k,T)$ by writing 
 \begin{equation}
w(k,T)=e^{\epsilon (k,T)/T}.
\label{dressE}
 \end{equation}
The point is that the grand partition function 
$Z_G$, corresponding to the thermodynamic 
potential (\ref{Omega}), is of the form of that 
for an ideal system of fermions with a 
complicated, $T$-dependent energy dispersion 
given by the dressed energy: 
\begin{equation}
Z_G=\prod_k(1+e^{-\epsilon (k,T)/T}).
\label{ptf}
\end{equation}
However, this fermion representation is not 
very useful, because of the implicit 
$T$-dependence of the dressed energy. To 
simplify, we consider the low-$T$ limit. By 
using the dressed energy, (\ref{forW}) reads
\begin{equation}
\epsilon(k,T)=k^2-\mu-T(1-\lambda)
\ln(1+e^{-\epsilon(k,T)/T}).
\end{equation}
Because there is no singularity in 
$\epsilon(k,T)$ at $T=0$, the zero 
temperature dressed energy is given by
\begin{equation}
\epsilon(k)=\Biggl\{ {\begin{array}{ll}
  (k^2-k^2_F)/\lambda , & |k|<k_F, \\
     \;\; k^2-k_F^2,&|k|>k_F.
 \end{array}}\label{EXP2}
\end{equation}
Denote
\begin{equation}
\epsilon(k,T)=\epsilon(k)
+\tilde{\epsilon}(k,T),
\label{EXP1}
\end{equation}
where 
\begin{equation}
\tilde{\epsilon}(k,0)=0.
\end{equation}
In the low-$T$ limit, one has
\begin{equation}
\epsilon(k,T)=\Biggl\{ 
{\begin{array}{ll}
\frac{k^2-\mu}{\lambda}-(\lambda^{-1}-1)T
\ln(1+e^{-|\epsilon(k)|/T}),&|k|<k_F,\\
(k^2-\mu)-(1-\lambda)T
\ln(1+e^{-|\epsilon(k)|/T}),&|k|>k_F,
\end{array}}.
\label{lowTd}
\end{equation}
Hence, 
\begin{equation}
\tilde{\epsilon}(k)=\Biggl\{ 
{\begin{array}{ll}
(1-\lambda^{-1})T\ln(1+e^{-|\epsilon(k)|/T}) , 
& |k|<k_F, \\
\;\;(\lambda-1)T\ln(1+e^{-|\epsilon(k)|/T}) ,
&|k|>k_F.
 \end{array}}
\label{EXP3}
\end{equation}
For low energies, one can consider only the 
excitations around the Fermi surface,
\begin{eqnarray}\displaystyle
\frac{\Omega(T)}{L}
&\approx&-\frac{T}{2\pi}\int_{-\infty}^{\infty}
dk \ln(1+e^{-\epsilon(k)/T})\nonumber\\
&+&\frac{1}{2\pi}\int_{-\infty}^\infty dk
\frac{ \tilde{\epsilon}(k,T)}
{1+e^{\epsilon(k)/T}}\nonumber\\
&\approx&\frac{1}{2\pi}\int_{-k_F}^{k_F}
dk \epsilon(k)-\frac{T}{\pi\lambda}
\int_{k_F-\delta}^{k_F}dk\ln(1
+e^{-|\epsilon(k)|/T})\nonumber\\
&&-\frac{T}{\pi}\int^{k_F+\delta}_{k_F}dk
\ln(1+e^{-|\epsilon(k)|/T}), 
\label{OmegaLT}
\end{eqnarray}
where the first term on the right hand side 
of the last equality is recognized as 
$\Omega(0)/L$. The cut-off $\delta$ is of 
order $O(T/v_s)$ (actually, a few times of $T/v_s$). 
Mathematically, we take the limit of $T\to 0$ 
followed by $\delta\to 0$. Using the integral 
formula
$$
\int_0^\infty dx \ln(1+e^{-x})=\frac{\pi^2}{12},
$$
we have the low-$T$ thermodynamic potential
\begin{equation}
\frac{\Omega(T)}{L}-\frac{\Omega(0)}{L}
=-\frac{\pi T^2}{6v_s},
\end{equation}
which implies that the theory is cut-off 
independent at low temperature.
Notice that $F=\Omega-\mu N$. Because we 
only consider the particle-hole excitations 
near the Fermi surface contribute to thermal 
excitations, $N(T)-N(0)=0$, which can be 
checked by an explicit calculation in terms 
of the definition of $\rho(k,T)$. Thus, we have
\begin{equation}
\frac{F(T)}{L}-\frac{F(0)}{L}
=\frac{\Omega(T)}{L}-\frac{\Omega(0)}{L}
=-\frac{\pi T^2}{6v_s}.\label{FREE}
\end{equation}
This means that the low energy behavior of 
the IEG is controlled by a $c=1$ CFT. This 
result can be verified by a finite-size 
scaling in the spatial direction,
\begin{equation}
\frac{F_L(0)}{L}-\frac{F(0)}{L}
=-\frac{\pi v_s}{6L^2},
\end{equation}
where $F_L(0)$ is the zero temperature free 
energy for a system with size $L$. (For 
details, see \cite{WuYu}.) 

The above relation agrees with the finite-size 
scaling of a conformally invariant system with 
central charge $c=1$. So we want to see whether 
the low-energy effective theory of the IEG is 
really a CFT. Let us start with the grand 
partition function (\ref{ptf}). At low 
temperature, the solution (\ref{lowTd}) 
leads to $\tilde\epsilon(k,T)=
O(Te^{-|\epsilon|/T})$, so one can simply 
replace $\epsilon(k,T)$ with $\epsilon(k)$ 
in the grand partition function:
\begin{equation}
Z_G\simeq \prod_k(1+e^{\beta\epsilon(k)}).
\end{equation}
Note that the dressed energy with $k$ 
outside the Fermi points $\pm k_F$ has a 
slope different from that with $k$ inside 
$\pm k_F$. The former is $\frac{2\pi}{L}$ 
and the latter $\frac{2\pi\lambda}{L}$. 
It is necessary to keep this in mind for 
writing down the correct ground state wave 
functions of the excluson system. Now that 
the dispersion $\epsilon(k)$ is $T$-independent, 
the grand partition function in the low-$T$ 
limit can be expressed in a fermionic 
representation as
\begin{equation}
Z_G={\rm Tr}e^{-\beta H_{\rm eff}},
\end{equation}
where the effective Hamiltonian is given by
\begin{equation}
H_{\rm eff}=\sum_k {\epsilon}(k) 
\;c_k^\dagger c_k,
\end{equation}
where $c_k^\dagger$ are fermionic creation 
operators. We also see that $\epsilon(k_F)=0$,
which can be used to define the Fermi momentum.

Physically, it is the phonon excitations that 
dominate the low-energy behavior of the system. 
In the low-$T$ limit, it is enough to consider 
the density fluctuations only near the Fermi 
points, $k\sim \pm k_F$, where the left- and 
right-moving sectors are separable and 
decoupled: 
\begin{equation}
H_{\rm eff}=H_{+} + H_{-}.
\end{equation}
Besides this, another important simplification 
for excitations near Fermi points in the low-$T$ 
limit is that their energy, $H_{\pm}$, has a 
linearized dispersion:
\begin{equation}
\epsilon_\pm (k)=\Biggl\{  {\begin{array}{ll}
\pm v_F(k\mp k_F), \;\;\; &|k|>k_F,\\ 
\pm v_F (k\mp k_F)/\lambda, &|k|<k_F.\\
\end{array}}
\label{LNH} 
\end{equation}
We note the `refractions' at $k=\pm k_F$, which 
implies to create a particle with pseudo-momentum 
$k$ and to create a hole with $k'$ cost different 
energies,  even if $|k-k_F|=|k'-k_F|$. The reason 
for this is that $k$ is not the actual momentum 
carried by $c_k^\dagger$, as we will see soon.

The key thing for bosonization is to construct 
a density fluctuation operator. Taking into 
account the different slopes for dressed
energy inside and outside the Fermi points, 
the density fluctuation operator at $k\sim k_F$ 
is constructed as follows:
\begin{eqnarray}
&&\rho_q^{(+)}=\displaystyle \sum_{k>k_F}
:c^\dagger_{k+q}c_k: 
+ \displaystyle \sum_{k<k_F-\lambda q}
:c^\dagger_{k+\lambda q}c_k:
\nonumber\\&&+ \displaystyle 
\sum_{k_F-\lambda q< k < k_F}
:c^\dagger_{\frac{k-k_F}{\lambda}+k_F+q}c_k: 
\label{cdensity}
\end{eqnarray}
for $q>0$. A similar density operator 
$\rho_q^{(-)}$ can also be defined at 
$k\sim -k_F$,
\begin{eqnarray}
&&\rho_q^{(-)}=\displaystyle 
\sum_{k<-k_F}:c^\dagger_{k-q}c_k: 
+ \displaystyle \sum_{k>-k_F+\lambda q}
:c^\dagger_{k-\lambda q}c_k:
\nonumber\\ && 
+ \displaystyle \sum_{-k_F+\lambda q> k >- k_F}
:c^\dagger_{\frac{k+k_F}{\lambda}-k_F-q}c_k: 
\end{eqnarray} 
To define the normal ordering we write, 
e.g.,
\begin{equation}
c_k=\Biggl\{{\begin{array}{ll}
c_k,&k>k_F,\\
d^\dagger_k,&k<k_F,\end{array}}
\end{equation}
where $d^\dagger_k$ is understood as a creation 
operator of a hole. Then normal ordering is done
as usual: putting the annihilation operators to
the right of the creation ones. Hence we have, 
e.g.,
\begin{eqnarray}
&&\rho_q^{(+)}=\displaystyle \sum_{k>k_F}
:c^\dagger_{k+q}c_k: 
+ \displaystyle \sum_{k<k_F-\lambda q}
:d_{k+\lambda q}d^\dagger_k:\nonumber\\ 
&&+ \displaystyle 
\sum_{k_F-\lambda q< k < k_F}
:c^\dagger_{\frac{k-k_F}{\lambda}+k_F+q}d^\dagger_k: 
\end{eqnarray}

Within the Tomonaga approximation \cite{comm1}, 
in which commutators are taken to be their 
ground-state expectation value, we obtain 
\begin{eqnarray}
&&[\rho_q^{(\pm)},\rho_{q'}^{(\pm)\dagger}]
\approx \langle 0|[\rho_q^{(\pm)},
\rho_q^{(\pm)\dagger}]|0\rangle\nonumber\\&&
=\sum_{k_F-\lambda q< k < k_F}
\langle 0|c_{k+\lambda q} 
c^\dagger_{k+\lambda q'}|0\rangle\nonumber\\
&&=\delta_{q, q'}\sum_{k_F-\lambda q< k < k_F}
1= \frac{ L}{2\pi} q\delta_{q,q'}
\label{density} 
\end{eqnarray}
Also, the commutators between $H_{\rm eff}$ 
and $\rho_q^{(\pm)}$ are
\begin{equation}
[H_{\pm},\rho_q^{(\pm)}]\approx
\langle0|[H_{\pm},\rho_q^{(\pm)}]|0\rangle
= \pm v_Fq\rho^{(\pm)}_q. 
\label{denHam}
\end{equation}
(\ref{density}) and (\ref{denHam}) describe 1-d 
free phonons with the sound velocity $v_{s}=v_F$ 
(so we have proved (\ref{VSF})). Introducing 
normalized boson annihilation operators 
\begin{equation}
b_q=\sqrt{2\pi/ qL}\,\rho_q^{(+)}, 
\hspace{.2in} \tilde{b}_q=\sqrt{2\pi/ qL}\,
{\rho}_q^{(-)\dagger}
\end{equation} 
and adding back the zero mode contributions, 
the bosonized Hamiltonian satisfying 
(\ref{density}) is given by  
\begin{equation}
H_B=v_s\{ \sum_{q>0}q(b_q^\dagger b_q
+\tilde{b}_q^\dagger \tilde{b}_q)
+\frac{1}{2}\frac{\pi}{L}[\lambda M^2
+\frac{1}{\lambda} J^2] \},
\label{bosonH}
\end{equation}
which agrees with the bosonized Hamiltonian 
(\ref{HmB}) in the Luttinger liquid theory.

In passing, we make a comment on linearization
of the dressed energy dispersion. When we did 
this, we changed the ground state energy, 
because we assumed that for all $k$ the 
spectrum is linear in $k$. However, we changed 
neither the ground state wave function, nor 
the low-$T$ physics. On the other hand, the 
linearized spectrum was valid only for phonon 
excitations, it has nothing to do with the
zero-mode excitations. So, after the linearized 
phonon part of the Hamiltonian is bosonized, we 
had to add back the zero-mode excitations.  

The construction of the bosonized momentum 
operator is a bit more tricky, because 
$c_k^\dagger$ does not carry a momentum $k$. 
Each term in (\ref{cdensity}) should carry 
the same momentum $q$, therefore the fermion 
created by $c_{k}^\dagger$ carries a dressed 
momentum $p$, which is related to $k$ by
\begin{eqnarray}
p(k)=\Biggl\{ {\begin{array}{lll}
 k-k_F+ (k_F/\lambda),\;\;\; & k>k_F, \\
 \;\; k/\lambda,  & |k|<k_F,.\\
 k+k_F- (k_F/\lambda), & k<-k_F.\\
\end{array}}
\label{dressP}
\end{eqnarray}
In terms of this variable, the linearized dressed 
energy $\epsilon (p)$ is of a simple form: 
$\epsilon_{\pm}(p)=\pm v_s(p\mp p_F)$, with
$p_F=k_F/\lambda$. The bosonized total momentum 
operator, corresponding to the fermionized 
$P=\sum_{k} p(k)\, c_{k}^{\dagger} c_{k}$, is 
\begin{equation}
P=\displaystyle\sum_{q>0}q(b_q^\dagger b_q-
\tilde{b}_q^\dagger \tilde{b}_q)
+ \pi (\bar{d}_0+M/L)\,J.\\
\label{bosonP}
\end{equation}

We see that the fundamental velocity relation, 
the bosonized Hamiltonian and momentum, and 
the selection rule of the quantum numbers in 
the Luttinger liquid theory can all be 
reproduced in IEG if we identify
\begin{equation}
\lambda\equiv e^{-2\varphi}.
\label{EQV}
\end{equation}
To say that IEG can be used to characterize 
the renormalization group {\it fixed points} of Luttinger liquids, 
we still need to check the conformal 
invariance of the bosonized theory of IEG, 
and to verify the critical properties of 
IEG reproduce those of the Luttinger liquids. 

\subsection{Effective Field Theory and 
Conformal Invariance}

To check conformal invariance, we need 
to rewrite the above bosonized effective 
Halmitonian (\ref{bosonH}) into a form 
of field theory in coordinate space. 
Employing the Fourier transformation, 
the density operator can be written as 
\begin{eqnarray}
\rho(x)&=&\rho_R(x)+\rho_L(x),\nonumber\\ 
\rho_R(x)&=&\frac{M_R}{L}+
\displaystyle\sum_{q>0}
\sqrt{\frac{q}{2\pi L\lambda}}
(e^{iqx}b_q+e^{-iqx}b_q^\dagger),\nonumber\\
\rho_L(x)&=&\displaystyle\frac{M_L}{L}+
\displaystyle\sum_{q>0}
\sqrt{\frac{q}{2\pi L\lambda}}
(e^{-iqx}\tilde{b}_q+e^{iqx}
\tilde{b}_q^\dagger),
\end{eqnarray}
where $M_{R,L}$ are given by $M=M_R+M_L$ 
and $\tilde{b}_q=b_{-q}$ for $q>0$.

The boson field $\phi(x)$, which is 
conjugated to $\rho(x)$ and satisfies 
\begin{equation}
[\phi(x),\rho(x')]=i\delta(x-x'),
\end{equation}
is given by 
\begin{eqnarray}
\phi(x)&=&\phi_R(x)+\phi_L(x),\nonumber\\
\phi_R(x)&=& \frac{\phi_0}{2}
+\frac{\pi J_Rx}{L}+i\displaystyle 
\sum_{q>0}\sqrt{\frac{\pi \lambda}{2qL}}
(e^{iqx}b_q-e^{-iqx}b_q^\dagger),\nonumber\\
\phi_L(x)&=& \frac{\phi_0}{2}
+\frac{\pi J_Lx}{L}+i\displaystyle 
\sum_{q>0}\sqrt{\frac{\pi \lambda}{2qL}}
(e^{-iqx}\tilde{b}_q-e^{iqx}\tilde{b}_q^\dagger),
\end{eqnarray}
with $J=J_R+J_L$. We have to assign the quantum 
numbers such that there are only two independent 
each other in $M_{R,L}$ and $J_{R,L}$. A 
consistent choice is
\begin{equation}
M_R=J_R,\hspace{.2in}M_L=-J_L.
\end{equation}
Then,
\begin{equation}
J=J_R+J_L,\hspace{.2in}M=J_R-J_L.
\end{equation}
Here $\phi_0$ is an angular variable
conjugated to $M$: $[\phi_0,M]=i$. 
The Hamiltonian (\ref{HmB}) becomes
\begin{equation}
H =\frac{1}{2}
\int_0^Ldx\; [\pi v_N\rho(x)^2
+v_J/\pi~(\partial_x\phi(x))^2],
\end{equation}
or by a field rescaling,
\begin{equation}
H =\frac{v_s}{2\pi}
\int_0^Ldx\; [\Pi(x)^2+(\partial_xX(x))^2],
\label{fieldH}
\end{equation}
where
\begin{equation}
\Pi(x)=\pi \lambda^{1/2}\rho(x),\hspace{.2in} 
X(x)=\lambda^{-1/2}\phi(x).
\end{equation}

With $X(x,t)=e^{iHt}X(x)e^{-iHt}$, the 
Lagrangian density reads 
\begin{equation}
{\cal L}=\frac{v_s}{2\pi}\,\partial_\alpha 
X(x,t)\,\partial^\alpha X(x,t).
\label{LGL}
\end{equation}
This is the Lagrangian density of a free 
scalar field theory in $1+1$-dimensions.
Writing the corresponding operators as the 
functionals of the scalar field $X(x,t)$, all 
correlation functions can be obtained by using 
the propagators of $X_R(x,t)$ and $X_L(x,t)$,
\begin{eqnarray}
\langle X_R(x,t)X_R(0,0)\rangle
=-\frac{1}{4}\ln(x-v_st),\nonumber\\
\langle X_L(x,t)X_L(0,0)\rangle
=-\frac{1}{4}\ln(x+v_st).
\label{prop}
\end{eqnarray}
The statistics of an operator in the 
theory can also be inferred by the 
commutators of the scalar fields,
\begin{equation}
[X_{R,L}(x),X_{R,L}(x')]=
\pm\frac{i\pi}{4}\theta(x-x').
\label{Comm}
\end{equation}

We recognize that $\cal{L}$ (\ref{LGL}) is 
the Lagrangian of a $c=1$ CFT\cite{CFT}, 
consistent with the finite-size scaling
(\ref{FREE}). Alternatively, it is easy to 
check that the theory is invariant under 
the conformal transformations generated 
by a set of the Virasoro generators
\begin{equation}
L_m=\frac{1}{2}\sum_{n=-\infty}^{\infty} 
\alpha_{n-m}\alpha_n, \hspace{.2in} 
\tilde{L}_m=\frac{1}{2}\sum_{n=-\infty}^{\infty} 
\tilde{\alpha}_{n-m}\tilde{\alpha}_n,
\end{equation}
where the oscillators $\alpha_m=m^{1/2}b_q$ 
and $\tilde{\alpha}_{-m}=m^{1/2}\tilde b_q^\dagger$ 
for $m=qL/2\pi>0$ being integers. $\alpha_0=
(\pi/2L)^{1/2}[J\lambda^{-1/2}-M\lambda^{1/2}]$ 
and $\tilde{\alpha}_0=
(\pi/2L)^{1/2}[J\lambda^{-1/2}+M\lambda^{1/2}]$.
The generators obey the Virasoro algebra with 
the central charge $c=1$,
\begin{equation}
[L^{\rm tot}_m,L^{\rm tot}_n]
=(m-n)L^{\rm tot}_{m+n}+\frac{1}{12}(m^3-m)
\delta_{m+n,0},
\end{equation}
with $L^{\rm tot}_m=L_m+\tilde{L}_m$.

Since $\phi_{0}$ is an angular variable, 
there is a hidden invariance in the theory 
under $\phi\to\phi+2\pi$. The field $X$ is 
thus said to be ``compactified'' on a circle, 
with a radius that is determined by the 
exclusion statistics \cite{WuYu,sm}: 
\begin{equation}
X\sim X+2\pi R,\;\;\; R^2=1/\lambda.
\label{INV}
\end{equation}

Noting the selection rule (\ref{select}), 
the Hamiltonian has a duality
\begin{equation}
\lambda\leftrightarrow 1/\lambda,
\hspace{.2in} M\leftrightarrow J, 
\end{equation}
which has referred to the particle-hole 
duality \cite{BerWu,NaWil}. Using the CFT 
terminology, this duality is represented 
as the duality of the compactified radii,
\begin{equation}
R\leftrightarrow 1/R.
\label{dua2}
\end{equation}
We note that this is different from the
duality relation $R\leftrightarrow 2/R$,
in the usual $c=1$ CFT \cite{CFT} 
compactifed on a circle. Actually, 
according to the standard terminology
in CFT \cite{CFT}, our selection rule
(\ref{select}) and duality relation 
(\ref{dua2}) make what we obtained above
a $c=1$ CFT compactified on an {\it orbifold}
$S^1/Z_2$, i.e., a circle folded by a 
reflection about a diameter, which 
topologically is a semi-circle or an interval.    
This difference can also be seen from the
grand partition function: Using the 
identification between $H_{\rm eff}$ and 
$L^{\rm tot}_0$, i.e., $H_{\rm eff}=v_s L^{\rm tot}_0$, 
the grand partition function of IEG (in the 
low-$T$ limit) can be rewritten as 
\begin{equation}
Z_G= Tr_{\cal H}[q^{L_0} \bar{q}^{\tilde{L}_0}],
\end{equation}
where $q=e^{iv_s\tau}$ with $\tau=i\beta=i/T$. 
Thus, the selection rule (\ref{select}) 
severely constrain the allowed values for 
the eigenvalues of $L_0$ and $\tilde{L}_0$. 
It makes the CFT we obtained have an unusual 
spectrum and duality relation, corresponding 
to the $c=1$ orbifold CFT \cite{CFT}. In next
subsection we will see that because of the
difference in the selection rules, the 
statistics of the allowed charge-1 operators 
in the two classes of CFT's are not the same.   

We note that a similar situation happens 
for the CFT that describes the low-$T$ 
behavior of the Calogero-Sutherland (C-S) 
model \cite{Cal,Suth}. This model has two 
different versions, with the long-range 
interactions being among bosons or among 
fermions, respectively. At low temperature, 
the two versions have different selection 
rules for the zero-mode quantum numbers, 
thus leading to different CFT's: The 
low-$T$ CFT for the bosonic C-S model is 
the usual $c=1$ CFT compactified on a 
circle, which has been studied extensively 
in the literatures \cite{KY,sm,ISO,ISR,CAR,WuYu}; 
while for the fermionic C-S model the low-$T$ 
limit gives rise to the $c=1$ orbifold CFT. 
This is because the selection rule for zero 
modes severely constrains the spectrum of 
the system, i.e., possible quantum numbers 
of the allowed excitations. (For details, 
see ref. \cite{WuYu}.) Thus, only the 
fermionic (not the bosonic) C-S model 
respects a duality relation $\lambda
\leftrightarrow 1/\lambda$ that coincides
with the particle-hole duality in IEG 
\cite{BerWu,NaWil}.

\subsection{Correlation Functions}

The CFT description of the IEG offers a better 
understanding for the space of quantum states 
in the theory.  States $V[X]|0\rangle$ or 
operators $V[X]$ are {\it allowed} only if 
they respect the invariance (\ref{INV}),
\begin{equation}
V[X+2\pi R]\equiv V[X], 
\end{equation}
with a given boundary condition restriction. 
Here, a Fermi or a Bose operator obeys the 
periodic boundary condition (PBC). So 
quantum numbers of quasiparticles are 
strongly constrained, in particular by the 
selection rule for zero-mode quantum numbers. 
For example, the primary fields obeying the 
PBC in the CFT are given by
\begin{eqnarray}
\phi_{M,J}(x)&\sim& f(J,X^0)
:e^{i(M\lambda^{1/2}+J/\lambda^{1/2})X_R(x)}\nonumber\\
&\times&e^{i(M\lambda^{1/2}-J/\lambda^{1/2})X_L(x)}:,
\nonumber\\
f(J,X^0)&=&e^{iJ(\lambda^{1/2}
-\lambda^{-1/2})X^0_R}
e^{-iJ(\lambda^{1/2}-\lambda^{-1/2})X^0_L}
\end{eqnarray}
where the prefactor $f(J,X^0)$ makes the fields 
satisfy the PBC, $M$ and $J$ eigenvalues of the 
number and current operators, and $X^0=\pi Mx/L$. 
The field carries the charge $M$ and current $J$. 
The conformal dimensions of the fields are
\begin{eqnarray}
h&=&\frac{1}{2}[(M\lambda^{1/2}+J/\lambda^{1/2})^2
+(M\lambda^{1/2}-J/\lambda^{1/2})^2]\nonumber\\
&=&M^2\lambda+J^2\lambda^{-1}.
\end{eqnarray}
The statistics of the field can be calculated 
by using (\ref{Comm}) and the statistics 
factors are 
\begin{eqnarray}
&&\exp\{i{\pi\over 4}[(M\lambda^{1/2}
+J/\lambda^{1/2})^2-(M\lambda^{1/2}
-J/\lambda^{1/2})^2]\}\nonumber\\
&=&(-1)^{MJ}.
\end{eqnarray}
Consider the charge-1 primary fields, with $M=1$. 
Therefore, they can only be fermions since $J=$ 
odd due to the selection rule.  The general 
charge-1 fermion operator is a linear combination 
of the charge-1 primary fields. A careful 
construction of the allowed fermion field with 
unit charge leads to 
\begin{eqnarray}
\Psi^\dagger_F(x,t)=\rho(x)^{1/2}
&&\sum_{m=-\infty}^{\infty} e^{iO_m}
:e^{i(\lambda^{1/2} + (2m+1)/\lambda^{1/2})X_{R}(x_-)}:
\nonumber \\
&& :e^{i(\lambda^{1/2}-(2m+1)/\lambda^{1/2})X_{L}(x_+)}:
\; ,
\label{fermiPsi}
\end{eqnarray}
where the prefactor $f$ has been suppressed and 
the hermitian, constant-valued operators $O_m$ 
satisfy \cite{comm2}
\begin{equation} 
[O_m, O_{m'}]=i\pi(m-m').
\end{equation}

The multi-sector density operator is the 
linear combination of those primary fields 
with $M=0$ and $J=$even,
\begin{eqnarray}
\hat{\rho}(x)&=&\Psi_F^\dagger(x)
\Psi_F(x)\nonumber \\
&=& \rho(x)\sum{}_m 
:\exp\{i2m[X_R(x)-X_L(x)]/\lambda^{1/2}\}:.
\end{eqnarray}

All the secondary fields in the CFT follow by 
considering the sound wave contribution to the 
conformal weight of the fields.

The correlation functions can easily be  
calculated by using the CFT techniques. For 
examples, the density-density and single
particle correlation functions are as follows,
\begin{eqnarray}
\langle \hat{\rho}(x,t)\hat{\rho}(0,0)\rangle 
&\approx&\bar{d}_0^2 \Biggl[1+\displaystyle
\frac{1}{(2\pi\bar{d}_0)^2\lambda}\Biggl(
\frac{1}{x_R^2}+\frac{1}{x_L^2}\Biggr)\nonumber \\
&+&{\displaystyle\sum_{m=1}^{\infty}} A_m
\frac{1}{[x_Rx_L]^{m^2/\lambda}}
\cos(2\pi\bar{d}_0mx)\Biggr],
\end{eqnarray}
and
\begin{eqnarray}
&&G(x,t)\equiv \langle \Psi^\dagger_F(x,t)
\Psi_F(0,0)\rangle \nonumber\\
\approx \bar{d}_0\displaystyle
\sum_{m=-\infty}^{\infty}
&&B_m\frac{1}{x_R^{(\lambda^{1/2}
+(2m+1)\lambda^{-1/2})^2/4}}
\nonumber\\&&
\frac{1}{x_L^{(\lambda^{1/2}
-(2m+1)\lambda^{-1/2})^2/4}}
\nonumber\\&&
e^{i(2\pi(m+\lambda/2)\bar{d}_0x+\mu t)},
\label{fgreen}
\end{eqnarray}
where $x_{R,L}=x\mp v_st$ and $A_m$ and 
$B_m$ regularization-dependent constants.

Usually a physical quantity, e.g., a boson 
field, satisfies the periodic boundary 
conditions (PBC). Hence, a charge-1 bosonic 
excitations are not allowed in the theory,
because it is anti-periodic. However, as we 
know, an anyon field needn't to obey the PBC. 
So in the theory, there may be allowed 
anyonic excitations. A charge-1 anyonic 
(or exclusonic) operator is a primary 
field that does not obey the PBC,
\begin{equation}
\Psi^\dagger_{\lambda}(x)=
:\Psi^\dagger_{F}(x)e^{i(\lambda^{1/2}
-\lambda^{-1/2})(X_R(x)-X_L(x))}:.
\end{equation}
The anyon commutation relation is easy to 
check: 
\begin{equation}
\Psi^\dagger_{\lambda}(x)
\Psi^\dagger_{\lambda}(x')
-e^{i\pi\lambda {\rm sgn}(x-x')}
\Psi^\dagger_{\lambda}(x')
\Psi^\dagger_{\lambda}(x)=0, \hspace{.2in}
{\rm for}~~x\not= x'.
\end{equation} 
In other words, the anyon field carries 
a fractional current. Or by the 
$M\leftrightarrow J$-duality, the anyon 
with integer $J$ carries a fractional
charge. The correlation function of the 
single-anyon reads \begin{eqnarray}
G(x,t;\lambda)&\equiv& \langle 
\Psi^\dagger_{\lambda}(x,t)
\Psi_{\lambda}(0,0)\rangle \\
\approx \bar{d}_0\displaystyle
\sum_{m=-\infty}^{\infty}
&B^a_m&\frac{1}{x_R^{(m+\lambda)^2/\lambda}}
\frac{1}{x_L^{m^2/\lambda}}
e^{i(2\pi(m+\lambda/2)x+\mu t)},
\label{green}
\end{eqnarray}
This correlation function coincides with 
the asymptotic one \cite{Ha} in the 
Calogero-Sutherland model. We see that 

\noindent (i) if $\lambda=1$, (\ref{green})
consists with (\ref{fgreen}); 

\noindent (ii)  there are no  boson 
excitations ($\lambda=0$) because $G(x,t;0)=0$; 

\noindent (iii) and moreover, $\lambda>0$ is 
implied since (\ref{green}) will diverge 
at the long distance if $\lambda<0$. 

\noindent (iv) Look at $m=0$. The critical 
exponents can be reads out,
$$
\eta_f=\lambda+\lambda^{-1}, \hspace{.2in} 
\eta_\lambda=2\lambda.
$$
Thus, 
$$
\begin{array}{ll}
\eta_f>\eta_\lambda,&{\rm if}~\lambda<1;\\
\eta_f<\eta_\lambda,&{\rm if}~\lambda>1.
\end{array}
$$

\noindent (v) The multi-sector density operator 
for exclusons is the same as that of the fermion. 

The single-hole state, i.e.\,  
$\Psi^\dagger_{1/\lambda} |0\rangle \equiv 
\Psi_\lambda (\lambda\to\lambda^{-1}) |0\rangle $, 
with charge $-1/\lambda$ alone 
is not allowed. The minimum allowed 
multi-hole state is given by 
$$
\Psi^\dagger_{1/\lambda}(x_1)...
\Psi^\dagger_{1/\lambda}(x_p)|0\rangle
$$
if $\lambda=p/q$ is rational. 
One may obtain, e.g.\ , 
$$
\langle [\Psi^\dagger_{1/\lambda}(x,t)]^p
[\Psi_{1/\lambda}(0,0)]^p\rangle
\sim [G(x,t;1/\lambda)]^p.
$$
A more interesting allowed operator is what 
creates $q$ particle excitations accompanied 
by $p$ hole excitations: 
$$
\hat{n}(x,t)=[\Psi^\dagger_{\lambda}(x,t)]^q
[\Psi^\dagger_{1/\lambda}(x,t)]^p.
$$
We note the similarity of this operator 
to Read's order parameter \cite{Read} for 
fractional quantum Hall fluids (in bulk). 
Its correlation function can be calculated 
by using Wick's theorem:
\begin{equation}
\langle \hat{n}(x,t)\hat{n}(0,0) \rangle\sim
[G(x,t;\lambda)]^q[G(x,t;1/\lambda)]^p.
\end{equation}
If the contribution from the $m=0$ sector 
dominates, then one gets 
$$\langle \hat{n}(x,t)\hat{n}(0,0) 
\rangle \sim (x-v_st)^{-(p+q)}$$.

\section{Two Extensions } 

Now we proceed to go beyond IEG. Two 
extensions will be discussed in this 
section: The one-component GIG with 
the mutual statistics, and the non-ideal 
gas with the Luttinger-type interactions.
In either case, we will show that the 
low-temperature behavior is that of an 
IEG, controlled by a single "effective 
statistics" parameter $\lambda_{eff}$,
whose value depends on the mutual 
statistics and the coupling constants
in the interactions. 

\subsection{Generalized ideal gas with 
mutual statistics} 

We turn to discussing the effects of mutual 
statistics. Consider a GIG with the statistics 
matrix (\ref{stat}) in momentum space given by 
\begin{equation}
g(k-k')=\delta(k-k')+\Phi(k-k').
\end{equation} Here
$\Phi(k)=\Phi(-k)$ is a smooth function. 
$\Phi(k-k')$ stands for mutual statistics 
between particles with different momenta;
for IEG $\Phi(k)=(\lambda-1)\delta (k)$. 
The thermodynamic properties of GIG is 
given by eq. (\ref{Omega}), but now 
$w(k,T)$ satisfies integral equation 
\cite{Wu,BerWu} which, in terms of the 
dressed energy (\ref{dressE}), is of the form 
\begin{equation}
\epsilon(k,T)=\epsilon_0(k)+T
\int^{\infty}_{-\infty} \frac{dk'}{2\pi}\;
\Phi (k-k')\,\ln (1+ e^{-\epsilon(k',T)/T})\; ,  
\end{equation}
where $\epsilon_0(k)\equiv k^{2}-\mu$. 
In the low-$T$ limit, it can be proven by 
the iteration \cite{WuYu} that
$\epsilon(k,T)=\epsilon(k)+O(T^2/v_s)$, 
where $\epsilon(k)$ is the zero-temperature 
dressed energy given below. At $T=0$, the 
Fermi momentum $k_F$ is determined by 
\begin{equation}
\epsilon(\pm k_F)=0.
\end{equation} 
Introduce 
\begin{eqnarray}
(\alpha\cdot\beta) [-k_{F},k_F] 
&\equiv& \int^{k_F}_{-k_F}
\frac{dk}{2\pi}\; \alpha (k)\, \beta (k)\, ,\\
(\Phi \cdot\alpha) (k;-k_{F},k_F]
&\equiv& \int_{-k_F}^{k_F} \frac{dk'}{2\pi}
\;\Phi (k-k') \,\alpha (k')\, .
\end{eqnarray}
Then both $\rho (k)$ and $\epsilon(k)$ 
in the ground state satisfy an integral 
equations like
\begin{equation}
\alpha (k) = \alpha_0(k)
-(\Phi\cdot \alpha )\, (k;-k_F,k_F]\;.
\label{inteq}
\end{equation}
The dressed momentum $p(k)$ is related 
to $\rho (k)$ by 
\begin{equation}
dp(k)=2\pi\rho (k)dk,\hspace{.2in} 
p(k)=-p(-k).\label{pk} 
\end{equation}
The ground state energy is given by 
\begin{equation}
E_0/L=(\epsilon_0\cdot \rho)[-k_F,k_{F}].
\end{equation}
Using the equation satisfied by $\rho(k)$, 
it can be expressed by the dressed energy
\begin{equation}
E_0/L =(\epsilon\cdot\rho_0)[-k_F,k_{F}].
\end{equation}

The above equations are of the same form 
as those in the thermodynamic Bethe ansatz 
\cite{YangYang}, hence the Luttinger-liquid 
relation \cite{Hald4},
$v_{s} = \sqrt{v_{N}v_{J}}$, remains true. 
A simple proof is sketched as follows. 
The sound velocity is well-known: 
\begin{equation}
v_s= \partial \epsilon(p_{F})/\partial p_{F}.
\end{equation} 
The charge velocity is given by 
\begin{equation}
v_N=v_s\, z(k_F)^{-2},
\end{equation}
where the dressed charge $z(k)$ \cite{Hald4}
is given by the solution to the integral equation 
\begin{equation}
z(k)=1- (\Phi \cdot z) (k;-k_F,k_{F}].
\end{equation}
This relation can be easily derived from the 
definitions 
\begin{eqnarray}
v_N&=&L \partial \mu/\partial N_{0},\nonumber\\ 
z(k)&=&-\delta \epsilon(k)/\delta \mu.
\end{eqnarray}
To create a persistent current, let us boost 
the Fermi sea by 
\begin{equation}
\pm k_{F} \to \pm k_F+\Delta,
\end{equation}
where $\Delta=z(k_F)/L\rho(k_F)$. Then the total 
energy of the state with the persistent current is 
\begin{eqnarray}
E_\Delta/L&=&(\epsilon_0\cdot\rho_\Delta)[-k_F
+\Delta,k_{F}+\Delta] \nonumber\\
&=& (\epsilon_\Delta\cdot\rho_0)
[-k_F+\Delta,k_{F}+\Delta],
\end{eqnarray} \label{515}
where
\begin{equation}
\rho_{\Delta} (k)=\rho_0(k)-
(\Phi\cdot \rho_\Delta)(k;-k_F+\Delta,k_{F}+\Delta]
\end{equation} and 
\begin{equation}
\epsilon_\Delta(k)=\epsilon_0(k)-
(\Phi \cdot\epsilon_\Delta)(k;-k_F+\Delta,k_{F}+\Delta].
\end{equation} 
Now, using the last expression for $E_\Delta$ and 
substituting $\epsilon_\Delta$ in (\ref{515}), we have
\begin{eqnarray}
&&E_\Delta/L=(\epsilon\cdot\rho_0)[-k_F,k_{F}]
+{\Delta^2\over 2}\epsilon'(k_F)\nonumber\\
&&\{\rho_0(k_F)+(\rho_0\cdot 2\pi F)(k_F;-k_F,k_F]\}
\nonumber\\
&&-{\Delta^2\over 2}\epsilon'(-k_F)\{\rho_0(-k_F)
+(\rho_0\cdot 2\pi F)(-k_F;-k_F,k_F]\}.
\end{eqnarray}
Here $F(k,k')$ is determined by 
\begin{equation}
F(k,k')=\frac{1}{2\pi}\Phi(k,k')
-\frac{1}{2\pi}\int_{-k_F}^{k_F}dk''\Phi(k,k'')F(k'',k').
\end{equation}
On the other hand, we note that the equation for
$\rho_0(k)$ can be rewritten as
\begin{equation}
\rho(k)=\rho_0(k)-(\rho_0\cdot2\pi F)(k;-k_F,k_F].
\end{equation}
Thus, we have
\begin{equation}
E_\Delta-E_0=L\Delta^2
\epsilon'(k_F)\rho (k_F)=(2\pi/L)v_s z(k_F)^{2}.
\end{equation} 
This verifies $v_J=v_s z(k_F)^2$. In view of 
eq. (\ref{Velo}), at low energies, the GIG 
looks like an IEG with 
\begin{equation}
\lambda_{eff}=z(k_{F})^{-2}.
\label{effstat}
\end{equation}

It can be shown that it is the effective statistics
(\ref{effstat}) that controls the low-$T$ critical 
properties of GIG, as $\lambda$ does for IEG\@. 
Linearization near the Fermi points and bosonization 
of the low-energy effective Hamiltonian go the same
way as before for IEG\@. The only difference now
is that the slope of the linearized dispersion for 
the dressed energy $\epsilon_\pm(k)=\pm \epsilon'(k_F)
(k\mp k_F)+\mu = \pm v_s(p(k)\mp p_F)+\mu$, 
is smooth at $k\sim \pm k_{F}$. So bosonization is 
standard and the bosonized Hamiltonian is the same 
as eq. (\ref{bosonH}) for IEG, only with $\lambda$ 
replaced by $\lambda_{eff}$. However, before going 
to the bosonization we need an effective Hamiltonian 
of the fermions with the dressed energy. Unlike the IEG, 
in the GIG case, $\epsilon(k,T)=\epsilon(k)+O(T^2/v_s)$. 
Now, we work out the $T$-expansion of $\epsilon(k,T)$ 
explicitly in the low-$T$ limit:
\begin{equation}
\epsilon(k,T)=\epsilon(k)+\tilde{\epsilon}(k,T)
+O(T^3/v_s^2).
\label{EXP}
\end{equation}
One finds that
\begin{equation}
\tilde{\epsilon}(k,T)=
\frac{\pi T^2}{6\epsilon'(k_F)}f(k),
\end{equation}
with the function $f$ determined by 
\begin{eqnarray}
f(k)&=&\Phi(k_F-k)-(\Phi\cdot f)(k;-k_F,k_F]\nonumber\\
&=&\Phi(k_F-k)-(\Phi\cdot \Phi)(k;-k_F,k_F]\nonumber\\ 
&&+(\Phi\cdot (\Phi\cdot \Phi)(k;-k_F,k_F]+...
\label{f}
\end{eqnarray}
Note that the equation that $\rho(k)$ obeys 
can be rewritten as
\begin{eqnarray}
&&\frac{\rho(k)}{\rho_0}=1-\int_{-k_F}^{k_F} dk' 
\{\Phi(k-k')\nonumber\\
&&+(\Phi\cdot\Phi)(k';-k_F,k_F]
(\Phi\cdot (\Phi\cdot\Phi)(k';-k_F,k_F]+...
\label{rr}
\end{eqnarray}
Integrating (\ref{f}) over $k$ and comparing 
with (\ref{rr}), one has
\begin{equation}
\int^{k_F}_{-k_F}\frac{dk}{2\pi} f(k)
=1-\frac{\rho(k_F)}{\rho_0},
\end{equation}
and then
\begin{equation}
\int^{k_F}_{-k_F}\frac{dk}{2\pi} 
\tilde{\epsilon}(k,T)=\frac{\pi T^2}
{6\epsilon'(k_F)}(1-2\pi\rho(k_F)).
\label{til}
\end{equation}

Substituting (\ref{EXP}) into the thermodynamic 
potential (\ref{Omega}), we have
\begin{equation}
{\Omega(T)\over L}=-\frac{T}{2\pi}
\int^\infty_{-\infty}dk \ln(1+e^{-\epsilon(k)/T})
+\frac{1}{2\pi}\int^\infty_{-\infty}
\frac{dk}{1+e^{\epsilon(k)/T}}\tilde{\epsilon}(k,T).
\label{OmegaCT}
\end{equation}
In the low-$T$ limit, the first term in the 
last equation gives
$$
{\Omega(0)\over L}-\frac{\pi T^2}{6\epsilon'(k_F)},
$$
with
$$
{\Omega(0)\over L}=\frac{1}{2\pi}
\int^{k_F}_{-k_F}dk\epsilon(k).
$$
and the second term is approximately given 
by (\ref{til}). Thus, 
\begin{equation}
{\Omega(T)\over L}-{\Omega(0)\over L}
=-\frac{\pi T^2}{6v_s},
\label{OmegaBA}
\end{equation}
which proves the central charge $c=1$ CFT 
behavior of the theory at the low energy. 

We may also confirm this from the finite size 
scaling in the spatial direction. To see this, 
we consider the discrete version of the equation 
in which the density $\rho_L(k_i)$ obeys
\begin{equation}
\rho_L(k_i)=\frac{1}{2\pi}
-\sum_{j\not=i}\Phi(k_i-k_j).
\end{equation}
Using the relation between discrete 
sum and integration
\begin{eqnarray}
&&\frac{1}{L}\sum_{n=N_1}^{N_2}f(\frac{I_n}{L})
=\int_{(N_1+1/2)/L}^{(N_2-1/2)/L}dx f(x)
\nonumber\\
&&+\frac{1}{24L^2}[f'((N_1-1/2)/L)
-f'((N_2+1/2)/L)]\nonumber\\ &&
+O(1/L^3),\label{FSS}
\end{eqnarray}
one has
\begin{eqnarray}
&&\rho_L(k)\approx\frac{1}{2\pi}
-(\Phi\cdot \rho_L)(k;-k_F,k_F]\nonumber\\
&&-\frac{1}{24L^2}\frac{1}{\rho(k_F)}
\Biggl[\frac{\partial \Phi(k-k')}{dk'}\Biggr]_{-k_F}
\nonumber\\ 
&&+\frac{1}{24L^2}\frac{1}{\rho(k_F)}
\Biggl[\frac{\partial \Phi(k-k')}{dk'}\Biggl]_{k_F}.
\end{eqnarray}
Denote 
\begin{equation}
\rho_L=\rho+\rho_1,
\end{equation}
where $\rho(k)$ is of the order $O(1/L^0)$ and 
$\rho_1(k)$ the order $O(1/L^2)$. Then, 
$\rho(k)$ is as defined and $\rho_1(k)$
is determined by
\begin{eqnarray}
&&\rho_1(k)=
-\frac{1}{24L^2}\frac{1}{\rho(k_F)}
\Biggl[\frac{\partial \Phi(k-k')}{dk'}
\Biggr]_{-k_F}\nonumber\\&&
+\frac{1}{24L^2}\frac{1}{\rho(k_F)}
\Biggl[\frac{\partial \Phi(k-k')}{dk'}
\Biggr]_{k_F}
-(\Phi\cdot \rho_1)(k;-k_F,k_F],
\label{rho1}
\end{eqnarray}
The corresponding thermodynamic potential reads
\begin{eqnarray}
&&\frac{\Omega_L(0)}{L}
=\frac{1}{L}\sum_i\epsilon_0(k(\frac{I_i}{L}))
\nonumber\\
&&=\int_{-k_F}^{k_F}dk \rho_L(k)\epsilon_0(k)
+\frac{1}{24L^2\rho(k_F)} [\epsilon_0'(k)|_{-k_F}-\epsilon_0'(k)|_{k_F}]\nonumber\\
&&=\int_{-k_F}^{k_F}dk \rho(k)\epsilon_0(k)
+\int_{-k_F}^{k_F}dk \rho_1(k)\epsilon_0(k)\nonumber\\
&&+\frac{1}{24L^2\rho(k_F)}
[\epsilon_0'(k)|_{-k_F}-\epsilon_0'(k)|_{k_F}].
\end{eqnarray}
The first term of the last equation is 
${\Omega(0)}/{L}$ and the rest, using 
(\ref{rho1}), can be written as
\begin{eqnarray}
&&-\frac{1}{24L^2\rho(k_F)}
\frac{\partial}{\partial k}\biggl(\epsilon_0(k)
+(-1)(\Phi\cdot \epsilon_0)\nonumber\\&&
+(-1)^2 ((\Phi\cdot\Phi)\cdot
\epsilon_0)+...)(k;-k_F,k_F]\biggr)_{k=k_F}\nonumber\\ 
&&+\frac{1}{24L^2\rho(k_F)}
\frac{\partial}{\partial k}\biggl(\epsilon_0(k)
+(-1)(\Phi\cdot \epsilon_0)\nonumber\\
&&+(-1)^2 ((\Phi\cdot\Phi)\cdot
\epsilon_0)+...)(k;-k_F,k_F]\biggr)_{k=-k_F}.
\end{eqnarray}
Recall the equation that $\epsilon(k)$ obeys, 
one has immediately,
\begin{equation}
\frac{\Omega_L(0)}{L}-\frac{\Omega(0)}{L}
=-\frac{\pi}{12L^2}\frac{\epsilon'(k)_{k_F}
-\epsilon'(k)_{-k_F}}{2\pi \rho(k_F)}
=-\frac{\pi v_s}{6L^2}.
\end{equation}
as desired.

Similar to the case of IEG, we also could 
have a fermion representation of the grand 
partition function with the 
temperature-dependent spectrum. To derive 
the low-energy effective theory, however, 
one rewrites the thermodynamic potential 
(\ref{OmegaCT}) in the low-$T$ limit as
\begin{equation}
\frac{\Omega(T)}{L}\approx
\frac{\Omega(0)}{L}-2T\rho(k_F)I(k_F,T),
\end{equation}
where 
\begin{eqnarray}
I(k_F,T)&=&\int_{k_F-\delta}^{k_F+\delta}dk 
\ln (1+e^{-|\epsilon(k)|/T})
\nonumber\\
&=&\int_{p_F-\delta}^{p_F+\delta}
\frac{dp}{2\pi}\rho(p)
 \ln (1+e^{-|\epsilon(k((p))|/T}).
\end{eqnarray}
That is, 
\begin{eqnarray}
\frac{\Omega(T)}{L}&=&\int_{-k_F}^{k_F}
\frac{dk}{2\pi}\epsilon(k)\nonumber
\\&-&\frac{T}{2\pi}\int_{-p_F+\delta}
^{p_F-\delta}dp\ln(1+e^{-|\epsilon(k(p))|/T}),
\end{eqnarray}
where $p$ is the physical (dressed) momentum. The 
grand partition function reads
\begin{equation}
Z_G\simeq \prod_{k'}(1+e^{-\beta\epsilon(k(k'))}),
\end{equation}
where $k'=k$ for $|k|<k_F-\delta$ and 
$k'=p$ for $|k|>k_F-\delta$. Now, we can 
have an effective Hamiltonian because 
${\epsilon(k)}$ is $T$-independent,
\begin{equation}
H_{\rm eff}=\sum_{k'} \epsilon(k(k'))
c_{k'}^\dagger c_{k'}. 
\label{Heff}
\end{equation}
Similar to the IEG case, the low-$T$ excitations 
can be considered by taking the linear 
approximation near the Fermi points, and after 
bosonization the zero-temperature excitations 
should be added back. The way to bosonize the 
linear Hamiltonian is also similar to the case 
of IEG. Because the dressed energy is smooth at 
the Fermi points now, the bosonization is even 
simpler. The density fluctuation operators are 
simply given by 
\begin{eqnarray}
\rho_q^{(+)}&=&\sum_{k\sim k_F}:c^\dagger_{p+q}c_p:
\nonumber\\
\rho_q^{(-)}&=&\sum_{k\sim -k_F}:c^\dagger_{p-q}c_p:.
\end{eqnarray}  
The commutators among $\rho^{(\pm)}_q$ and $H_\pm$ 
are
\begin{eqnarray}
[\rho_q^{(\pm)},\rho_{q'}^{(\pm)\dagger}]
&\approx& \langle 0|[\rho_q^{(\pm)},
\rho_q^{(\pm)\dagger}]|0\rangle
=\sum_{p_F-q< p < p_F}\langle 0|c_{p+q}
c^\dagger_{p+ q'}|0\rangle\nonumber\\
&=&\delta_{q, q'}\sum_{p_F-q< p < p_F}1=
\frac{L}{2\pi} q\delta_{q,q'} 
\end{eqnarray}
and
\begin{equation}
[H_{\pm},\rho_q^{(\pm)}]\approx
\langle0|[H_{\pm},\rho_q^{(\pm)}]|0\rangle
= \pm v_Fq\rho^{(\pm)}_q. 
\label{HHH}
\end{equation}
Introducing the normalized bosonic
annihilation operators 
\begin{equation}
b_q=\sqrt{2\pi/ qL}\,\rho_q^{(+)}, 
\hspace{.2in} \tilde{b}_q=\sqrt{2\pi/ qL}\,
{\rho}_q^{(-)\dagger}
\end{equation}
and adding back the zero-mode contributions, 
the bosonized Hamiltonian satisfying (\ref{HHH}) 
is given by  
\begin{equation}
H_B=v_s\{ \sum_{q>0}q(b_q^\dagger b_q
+\tilde{b}_q^\dagger \tilde{b}_q)
+\frac{1}{2}\frac{\pi}{L}[\lambda_{\rm eff} M^2
+\frac{1}{\lambda_{\rm eff}} J^2] \},
\label{bosonH1}
\end{equation}
which agrees with the bosonized Hamiltonian 
(\ref{HmB}) in the Luttinger liquid theory. 
We see that with $\lambda$ replaced by 
$\lambda_{\rm eff}$, the bosonized Hamiltonian 
for the GIG is the same as that for the IEG. 
So, all consequences we have obtained from the 
bosonized Hamiltonian in the IEG case can be 
applied to the GIG case. Especially, there is 
an (allowed) $\Psi_{\lambda_{eff}}^\dagger$ 
describing the particle excitation near the 
Fermi surface with both anyon and exclusion 
statistics being $\lambda_{eff}$. In this sense, 
one may say that the effect of mutual statistics 
is to renormalize the statistics matrix.

Here we remark that in IEG,
$\Phi(k,k')=(\lambda-1)\,\delta(k-k')$ is not 
smooth, so the dressed charge has a jump at 
$k_F$: $z(k_F^+)=1$ and $z(k_F^-)=\lambda^{-1}$ 
for $k_F^\pm=k_F\pm 0^+$. The general 
Luttinger-liquid relation is of the form 
\begin{equation}
v_N=v_s[z(k_F^+)z(k_F^-)]^{-1},
~~~v_J=v_sz(k_F^+)z(k_F^-).
\end{equation} 

\subsection{Non-ideal Gas}

Finally, we examine non-ideal gases, e.g., with 
general Luttinger-type density-density interactions,  
\begin{eqnarray}
H&=&H_{\rm eff}+H_I,\nonumber\\
H_I&=&\displaystyle\frac{\pi}{L}
\sum_{q\geq 0}[U_{q}(\rho_q\rho_q^\dagger
+\tilde{\rho}_q\tilde{\rho}_q^\dagger)
+ V_{q}(\rho_q\tilde{\rho}_q^\dagger
+\tilde{\rho}_q\rho_q^\dagger)],
\end{eqnarray}
where $H_{\rm eff}$ is given by  
(\ref{Heff}) describing a GIG, and $\rho_q$ 
and $\tilde{\rho}_q$ are the excluson density 
fluctuations near $\pm k_F$ respectively. After 
bosonization, the total Hamiltonian remains 
bilinear in densities:
\begin{eqnarray}
H&=&H_B+H_I\nonumber\\
&=& \frac{1}{2}\sum_{q>0}q[(v_s+U_q)(b_q^\dagger b_q
+\tilde{b}_q^\dagger \tilde{b}_q+b_qb_q^\dagger
+\tilde{b}_q\tilde{b}_q^\dagger) \nonumber\\
&&+V_q(b_q^\dagger \tilde{b}^\dagger_q
+b_q \tilde{b}_q+\tilde{b}^\dagger_qb_q^\dagger
+\tilde{b}_qb_q)] +\frac{1}{2}\frac{\pi}{L}
[v_N M^2+v_J J^2]\nonumber\\
&&+\frac{\pi}{L}[U_0(M_R^2+M_L^2)+2V_0M_RM_L]-
\sum_{q>0}v_sq
\end{eqnarray}
Using the Bogoliubov transformation, the 
Hamiltonian can be easily diagonalized
\begin{eqnarray}
H&=&\sum_{q>0}\omega_q(a_q^\dagger a_q
+\tilde{a}_q^\dagger\tilde{a}_q)
+\frac{1}{2}(\pi/L)[\tilde{v}_NM^2
+\tilde{v}_JJ^2]+{\cal E}_0,\nonumber\\
{\cal E}_0&=&\sum_{q>0}(\omega_q-v_sq),
\end{eqnarray}
where
\begin{eqnarray}
a_q^\dagger&=&\cosh~\tilde{\varphi}_0
~b^\dagger_q-\sinh~\tilde{\varphi}_0~
\tilde{b}_q^\dagger,\nonumber\\
\tilde{a}_q^\dagger&=&\cosh~\tilde{\varphi}_0
~\tilde{b}^\dagger_q
-\sinh~\tilde{\varphi}_0~b_q^\dagger,
\end{eqnarray}
and the renormalized velocities are
\begin{eqnarray} 
v_{s}&\to&\tilde{v}_s=
|(v_s+U_0)^2-V_0^2|^{1/2},\nonumber\\
v_{N}&\to&\tilde{v}_N=
\tilde{v}_s e^{-2\tilde{\varphi}_0},
\nonumber\\
v_{J}&\to&\tilde{v}_J=\tilde{v}_s
e^{2\tilde{\varphi}_0}.
\end{eqnarray} 
with the controlling parameter 
$\tilde{\varphi}_0$ determined by
\begin{equation}
\tanh(2\tilde{\varphi}_0)
=\frac{v_J-v_N-2V_0}{v_J+v_N+2U_0}.
\end{equation}
Thus, the Luttinger-liquid relation 
((\ref{Velo}) survives with $\lambda_{eff}$ 
of GIG renormalized to
\begin{equation}
\tilde{\lambda}_{eff} = e^{-2\tilde{\varphi_0}}.
\end{equation}
Note that the new fixed point depends both 
on the position of the Fermi points and 
on the interaction parameters $U_{0}$ and $V_{0}$, 
leading to ``non-universal''exponents.

\section{Discussions and conclusions}

In conclusion, we have shown that 1-d IEG 
(without mutual statistics) exactly reproduces
the low-energy and low-$T$ properties of 
(one-component) Luttinger liquids. This gives 
rise to the following physical picture: At 
low temperature, the Luttinger liquids can be 
approximately thought of as an IEG consisting 
of quasiparticle excitations. Introducing mutual 
statistics or/ and Luttinger-type interactions 
among these excitations only shifts the value of 
$\lambda_{eff}$. Thus the essence of Luttinger 
liquids is to have an IEG obeying FES as their 
fixed point. This is our characterization of 
Luttinger liquids in terms of FES. 

In this way, we have explicitly answered the 
three questions raised in the introduction 
about Luttinger liquids:
\begin{itemize}
\item The physical meaning of the Haldane's
controlling parameter is the quasiparticle's
effective statistics, $\lambda_{eff}$. 
\item The Luttinger liquids, more precisely,
the IEG, indeed describe the infrared (or 
low-energy) fixed points in 1-d systems, 
since their effective field theory at low
energy is conformally invariant. However,
these fixed points are {\it not} isolated; 
they form a fixed-point line. Both the 
chemical potential and coupling constants
are relevant perturbations that can drive
the fixed point to move along the line,
corresponding to the "renormalization" of
the effective statistics $\lambda_{eff}$
and leading to "non-universal" exponents.
\item It is conceivable that some strongly 
correlated systems, exhibiting non-Fermi 
liquid behavior, in two or higher dimensions 
may also be characterized as having a GIG 
with appropriate statistics matrix as their 
low-energy or low-temperature fixed point.
This is because the concept of exclusion
statistics is independent of spatial 
dimeniosnality of the system.  
\end{itemize}

Moreover, we also showed that the effective 
field theory of 1-d IEG is a CFT with central 
charge $c=1$ and compactified radius 
$R=\sqrt{1/\lambda}$. The particle-hole duality 
of the exclusons implies the CFT has an unusual 
duality $R\leftrightarrow 1/R$, meaning that
the CFT belongs to a new variant of the $c=1$ 
CFT's, i.e the ones that are compactified on an 
"orbifold" $S^1/Z_2$ rather than on a circle. 
Physically, the differences are due to different 
constraints on the zero-mode quantum numbers. 
The CFT explanation makes a better understanding 
of the single-particle operators, especially, 
the anyonic (or exclusonic ) ones. Also, the CFT 
techniques provide a systematic way to calculate 
the correlation functions. 

Finally we observe several additional 
implications of this work: 1) Our bosonization 
and operator derivation of CFT at low energies 
or in low-$T$ limit can be applied to Bethe 
ansatz solvable models, including the long-range 
(e.g., Calogero-Sutherland) one \cite{WuYu}. 
2) Here we have only consider one-species cases, 
i.e., with excitations having no internal 
quantum numbers such as spin. Our bosonization 
and characterization of Luttinger liquids
are generalizable to GIG with multi-species, 
with the effective statistics matrix related to 
the dressed charge matrix \cite{WuYu}. 
3) The chiral current algebra in eqs. 
(\ref{density}) and (\ref{denHam})with 
$\lambda=1/m$ coincides with that derived 
by Wen\cite{Wen} for edge states in $\nu=1/m$ 
fractional quantum Hall fluids. So these edge 
states and their chiral Luttinger-liquid fixed 
points can be described in terms of chiral IEG\@.

This work was supported in part by the U.S. NSF 
grant PHY-9309458, PHY-9970701 and NSF of China.

\appendix

\section{The harmonic fluid description}

In coordinate space, there is a harmonic fluid 
description \cite{Hald3} of the Luttinger liquid. 
Instead of the $\theta$-$\phi$ representation 
that Haldane originally used, we prefer the 
right-left-moving representation. The density 
operator can be written as the Fourier 
transformations
\begin{eqnarray}
\rho(x)&=&\rho_R(x)+\rho_L(x),\nonumber\\ 
\rho_R(x)&=&\frac{M_R}{L}+\displaystyle
\sum_{q>0}\sqrt{\frac{q}{2\pi L e^{-2\varphi}}}
(e^{iqx}b_q+e^{-iqx}b_q^\dagger),\nonumber\\
\rho_L(x)&=&\displaystyle\frac{M_L}{L}+
\displaystyle\sum_{q>0}
\sqrt{\frac{q}{2\pi Le^{-2\varphi}}}
(e^{-iqx}\tilde{b}_q+e^{iqx}
\tilde{b}_q^\dagger),
\end{eqnarray}
where $M_{R,L}$ are given by $M=M_R+M_L$ and  
$\tilde{b}_q=b_{-q}$ for $q>0$.

The boson field $\phi(x)$, which is conjugated 
to $\rho(x)$ and satisfies 
\begin{equation}
[\phi(x),\rho(x')]=i\delta(x-x'),
\end{equation}
is given by 
\begin{eqnarray}
&&\phi(x)=\phi_R(x)+\phi_L(x),\nonumber\\
&&\phi_R(x)= \frac{\phi_0}{2}+\frac{\pi J_Rx}{L}+i
\displaystyle \sum_{q>0}\sqrt{\frac{\pi e^{-2\varphi}}
{2qL}}(e^{iqx}b_q-e^{-iqx}b_q^\dagger),\nonumber\\
&&\phi_L(x)= \frac{\phi_0}{2}+\frac{\pi J_Lx}{L}+i
\displaystyle \sum_{q>0}\sqrt{\frac{\pi e^{-2\varphi}}
{2qL}}(e^{-iqx}\tilde{b}_q-e^{iqx}\tilde{b}_q^\dagger),
\nonumber\\
&&
\end{eqnarray}
with $J=J_R+J_L$. We have to assign the quantum 
numbers such that there are only two independent
variables in $M_{R,L}$ and $J_{R,L}$. A consistent 
choice is
\begin{equation}
M_R=J_R,\hspace{.2in}M_L=-J_L.
\end{equation}
Then,
\begin{equation}
J=J_R+J_L,\hspace{.2in}M=J_R-J_L.
\end{equation}
Here $\phi_0$ is an angular variable
conjugated to $M$: $[\phi_0,M]=i$. 
The Hamiltonian (\ref{HmB}) becomes
\begin{equation}
H =\frac{1}{2}
\int_0^Ldx\; [\pi v_N\rho(x)^2
+v_J/\pi~(\partial_x\phi(x))^2],
\end{equation}
or by a field rescaling,
\begin{equation}
H =\frac{v_s}{2\pi}
\int_0^Ldx\; [\Pi(x)^2+(\partial_xX(x))^2],
\label{fieldH1}
\end{equation}
where
\begin{equation}
\Pi(x)=\pi e^{-\varphi}\rho(x),\hspace{.2in} 
X(x)=e^\varphi\phi(x).
\end{equation}

With $X(x,t)=e^{iHt}X(x)e^{-iHt}$, the 
Lagrangian density reads 
\begin{equation}
{\cal L}=\frac{v_s}{2\pi}\,\partial_\alpha 
X(x,t)\,\partial^\alpha X(x,t),
\label{LGL1}
\end{equation}
which describes a free scalar field theory 
in $1+1$-dimensions.


\begin{references}
\bibitem{comm0} It is easy to understand why 
it is $\delta F_0$ rather than $\delta E_0$ 
that exactly contains the zero-mode 
contributions in (\ref{HmB}): In writing
down the Hamiltonian (\ref{HmB}), we have 
shifted the zero point of the energy to 
the Fermi surface.
\bibitem{comm1} Alternatively, we may extend 
the Fermi sea to that for two separate right- 
and left-moving fermions with linear dispersion 
for $k$ covering the whole real axis in each 
sector. Then our commutation relations are 
exact. (See \cite{ML})
\end{references}
\end{document}